\documentclass[aps,prl,twocolumn,groupedaddress,showpacs]{revtex4}

\usepackage{graphicx}
\usepackage{amsmath}

\begin{document}

%\preprint{}

\title{Multiple domain formation induced by modulation instability in two-component Bose-Einstein condensates}


\author{Kenichi Kasamatsu and Makoto Tsubota}
%\thanks{}

\affiliation{
Department of Physics,
Osaka City University, Sumiyoshi-Ku, Osaka 558-8585, Japan}


\date{\today}

\begin{abstract}
The dynamics of multiple domain formation caused by the modulation instability of two-component Bose-Einstein condensates in an axially symmetric trap are studied by numerically integrating the coupled Gross-Pitaevskii equations. The modulation instability induced by the intercomponent mean-field coupling occurs in the out-of-phase fluctuation of the wave function and leads to the formation of multiple domains that alternate from one domain to another, where the phase of one component jumps across the density dips where the domains of the other exist. This behavior is analogous to a soliton train, which explains the origin of the long lifetime of the spin domains observed by Miesner {\it et al.} [Phys. Rev. Lett. {\bf 82}, 2228 (1999)].
\end{abstract}

%\keywords{}

\maketitle
Spatial pattern formation in systems that are attempting to reach equilibrium is one of the most important problems in nonlinear and nonequilibrium physics, and is encountered in many fields of physical science \cite{Cross}. The modulation instability (MI) is an indispensable mechanism for understanding pattern formation from a uniform system. MI is a general phenomenon in nonlinear wave equations, in which weak spatial perturbations with a certain range of wave numbers grow exponentially into a train of localized waves as a result of an interplay between nonlinearity and diffraction effects \cite{Agrawalbook}. 

Recently, the MI has been discussed in relation to the dynamics of ultra-cold atomic-gas Bose-Einstein condensates (BECs). This interest was triggered by the experimental observation of collapsing dynamics \cite{Donley} and the generation of matter-wave soliton trains \cite{Strecker} in BECs with an attractive interaction. The atom-atom interaction introduces nonlinearity into the system. The key parameter of this interaction is the s-wave scattering length, which can be controlled experimentally by the Feshbach resonance \cite{Inouye}. Solutions of these dynamics need nonlinear analysis beyond the MI; some theoretical and numerical studies have reported pattern formation caused by the MI in an attractive one-component BEC \cite{Saito,Kawaja,Salasnich,Kamchatnov,Carr}. Whereas these studies are concerned with the instability of self-wave modulation under attractive nonlinearity, if there are more than two wave fields, the coupling between multicomponent waves, called cross-phase modulation in the field of nonlinear optics, can also gives rise to MI even if all interactions are repulsive \cite{Agrawal,Goldstein}. The present study treats the nonlinear dynamics of pattern formation caused by cross-phase MI in two-component BECs by numerical simulation of the coupled Gross-Pitaevskii (GP) equations. 

Pattern formation in a multicomponent BEC was observed by Miesner {\it et al.} \cite{Miesner}. Miesner {\it et al.} first prepared a BEC of $^{23}$Na atoms with the hyperfine state $|F=1,m_{F}=1 \rangle = | 1 \rangle$ in a cigar-shaped optical trap, and then placed half of the atoms of $| 1 \rangle$ in the $|F=1,m_{F}=0 \rangle = | 2 \rangle$ state non-adiabatically by using rf irradiation.  Then, they made use of the quadratic Zeeman effect by applying a magnetic field to prevent the $|F=1,m_{F}=-1 \rangle$ component from appearing. As a result, the system could be regarded as binary. Within 100 msec after the placement, they observed the multiple domains that alternated from one domain to another. These domains had a typical sizes of 40 $\mu$m and had a very long lifetime of about 20 seconds in the absence of a magnetic field gradient. 

Two approaches have been used to gain an understanding of these long-lived domains. First, mean-field theory predicts that phase separation occurs in the equilibrium state of two-component BECs when the scattering lengths satisfy the relation $a_{12}>\sqrt{a_{1}a_{2}}$ \cite{Tim}, where $a_{1}$ and $a_{2}$ are the s-wave scattering lengths for same-species collisions, and $a_{12}$ for different-species ones. Because $a_{1} = a_{12} = 2.75$nm and $a_{2}=2.65$nm for $^{23}$Na atoms \cite{Stenger,Burke}, the condition $a_{12} > \sqrt{a_{1}a_{2}}$ is met and the condensates tend to separate spatially. Second, some authors pointed out that the multiple domain formation is related to MI due to the appearance of imaginary frequencies in the Bogoliubov modes of two miscible condensates \cite{Mueller,Ao2} and analyzed partially the nonlinear dynamics after this instability by numerical simulations \cite{Pu,Robins,Chui}. These findings are useful for identifying the conditions for multiple domains and some of the qualitative features of the dynamics, but there is much more to be learned about the nontrivial dynamical features of this sytem and the observation in Ref. \cite{Miesner}.

This work resolves the above problem by numerically solving the coupled GP equations that describe the time evolution of two-component BECs at very low temperatures: 
\begin{equation}
i\hbar \frac{\partial \psi_{i}}{\partial t} = \biggl[ -\frac{\hbar^{2} \nabla^{2}}{2m}
+V_{i} -\mu_{i} + g_{i} |\psi_{i}|^{2} + g_{12} |\psi_{3-i}|^{2} \biggr] \psi_{i}
 \label{2tgpe}
\end{equation}
with $i=1$ or $2$, where $\psi_{i}({\bf r}, t)$ is the condensate wave function in the hyperfine state $| 1 \rangle$ and $| 2 \rangle$, $m$ the atomic mass and $\mu_{i}$ the chemical potential. The atom-atom interactions are defined as $g_{i}=4 \pi \hbar^{2} a_{i}/m$ for intracomponent and $g_{12}=4 \pi \hbar^{2} a_{12}/m$ for intercomponent, corresponding to self-phase modulation and cross-phase modulation in nonlinear optics, respectively. We assume axial symmetry $(x,y,z) \rightarrow (r,z)$ and use an axisymmetric harmonic potential $V_{i}=m (\omega_{\perp}^{2} r^{2}+\omega_{z}^{2} z^{2})/2$. Following the condition of Ref.\cite{Miesner}, we use the mass of $^{23}$Na atoms $m=3.81 \times 10^{-26}$ kg, the trapping frequency $\omega_{\perp}/2\pi=500$ Hz and the aspect ratio $\lambda=\omega_{z}/\omega_{\perp}=1/70$. The wave function is normalized by the particle number as $\int d {\bf r} |\psi_{i}|^{2} =N_{i}$.

To show that the MI is caused by the cross-phase modulation, we first consider the stability of miscible two-component BECs in a homogeneous system \cite{Goldstein,Mueller,Ao2}, where the equilibrium densities satisfy the relation $g_{i}n_{i0}+g_{12}n_{3-i0}=\mu_{i}$. When each wave function is assumed to be written as $\psi_{i}({\bf r}, t)=\sqrt{n_{i0}} + A_{i} e^{i({\bf k} \cdot {\bf r} - \Omega t)}$ with the complex-valued amplitude $A_{i}$, the linearized equation with respect to the fluctuation provides a set of equations for the amplitude and gives the dispersion relation between the frequency $\Omega$ and the wave number $k$ 
\begin{equation}
\Omega_{\pm}^{2} = \frac{\hbar^{2} k^{2}}{2 m} \biggl( \frac{\hbar^{2} k^{2}}{2m} + g_{1}n_{10} + g_{2}n_{20} \pm \sqrt{\Delta} \biggr) \label{modufreq}
\end{equation}
with $\Delta=(g_{1}n_{10} - g_{2}n_{20})^{2} +4 g_{12}^{2} n_{10}n_{20}$. The condition of the MI depends on both the sign and the magnitude of the coupling constants. In this work, we focus on the case in which the signs of all coupling constants are positive. Other cases will be discussed elsewhere. In this case, if 
\begin{equation}
4g_{12}^{2}n_{10}n_{20} > \biggl( \frac{\hbar^{2}k^{2}}{2m} + 2g_{1} n_{10} \biggr) 
\biggl( \frac{\hbar^{2}k^{2}}{2m} + 2g_{2} n_{20} \biggr) \label{instineq}
\end{equation}
is satisfied, the branch $\Omega_{-}^{2}$ becomes negative, which implies an appearance of purely imaginary frequencies in the excitation modes and thus the miscible condensates are unstable. Equation (\ref{instineq}) gives the necessary condition $g_{12}^{2}>g_{1}g_{2}$ for an instability to occur in the range of wave number $0<k<k_{c} \equiv [2m ( \sqrt{\Delta}-g_{1}n_{10}-g_{2}n_{20})/\hbar^{2}]^{1/2}$. This instability is of particular importance because it occurs even for condensates with repulsive interactions, in contrast MI in a homogeneous single-component condensate occurs for only attractive interactions \cite{Agrawal}. 

To see the dynamics caused by the MI, we did numerical simulations of the coupled GP Eq. (\ref{2tgpe}). The numerical method used for solving the time-dependent GP equations was the Crank-Nicholson implicit scheme. As an initial state, we prepared the stationary solution consisting of only the $| 1 \rangle$ state by solving $(-\hbar^{2}\nabla^{2}/2m+V_{1}+g_{1}|\psi_{0}|^{2})\psi_{0}=\mu_{1}\psi_{0}$, where we set $N=\int d {\bf r} |\psi_{0}|^{2} = 2 \times 10^{6}$. Because the condensate is trapped in a highly elongated potential with $\lambda=1/70$, the axial direction satisfies the Thomas-Fermi limit $\mu_{1}/\hbar \omega_{z} \simeq 610 \gg 1$. In contrast, the radial direction does not satisfy this limit because $\mu_{1}/\hbar \omega_{\perp} \simeq 8.7 \sim {\cal O}(1)$. In the experiment, an rf pulse was applied instantaneously to transfer half of component $| 1 \rangle$ to component $| 2 \rangle$. In our simulation, we used the same wave function $\psi_{0}$ for both initial states $| 1 \rangle$ and $| 2 \rangle$, but their particle numbers are half of the original one, that is, $N_{1}=N_{2}=1 \times 10^{6}$. The sudden turn on of the intercomponent coupling $a_{12}$ and the mismatch between $a_{1}$ and $a_{2}$ of the initial configuration result in perturbations of the condensates. 

\begin{figure}
\includegraphics[height=0.40\textheight]{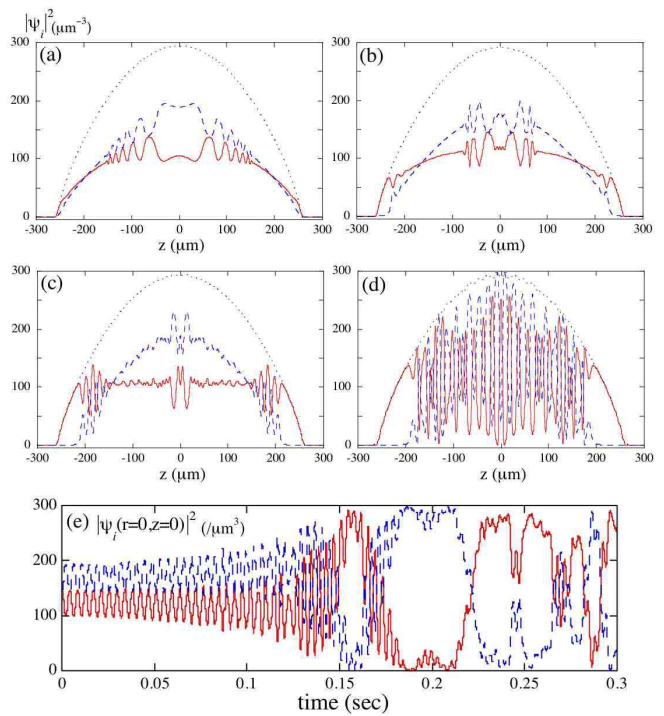}
\caption{Time development of the condensate densities $|\psi_{1}|^{2}$ (solid-curve), $|\psi_{2}|^{2}$ (dashed-curve) and total density $n_{T}=|\psi_{1}|^{2}+|\psi_{2}|^{2}$ (dotted-curve) along the $z$-axis at $r=0$ for $N_{1}=N_{2}=10^{6}$, $a_{1}=a_{12}=2.75$nm, $a_{2}=2.65$nm and $\lambda=1/70$. Time of each figure is (a) $t=40$ msec,  (b) $t=80$ msec,  (c) $t=120$ msec and (d) $t=160$ msec. The spatial grid is $[N_{r},N_{z}]=[50,4400]$ with the mesh size $\Delta_{r}=\Delta_{z}=0.14$ $\mu$m. Figures (e) shows time development of condensate densities at $r=0$ and $z=0$.}
\label{domedome}
\end{figure}
Figure \ref{domedome} shows the dynamics of multiple domain formation for $a_{1}=a_{12}=2.75$nm and $a_{2}=2.65$nm. As soon as the simulation began, rapidly oscillating density ripples were excited at the central region as shown in Fig. \ref{domedome}(a), propagating along the $z$-axis. The density wave is out-of-phase between the two components, so that the total density $n_{T}=|\psi_{1}|^{2}+|\psi_{2}|^{2}$ keeps its initial shape. This oscillation comes from the radial degree of freedom, which is tightly confined so that the mode neither grows nor does it form a multi-domain [see also Fig. \ref{domedome}(e)]. As time passes, because $a_{1}>a_{2}$, the axial width of $|\psi_{1}|^{2}$ expands rather than $\psi_{2}$ as seen in Fig. \ref{domedome}(b). Then, the amplitude of the density waves grows first near the edge of the condensate [Fig. \ref{domedome}(b) and (c)], eventually developing into multiple condensate domains as shown in Fig. \ref{domedome}(d). The domains of two components locate alternately and the total density does not change even after the domain formation. The domain formation starts at about $120$ msec and the size of one domain is about 20 $\mu$m, in good agreement with the experimental observation \cite{Miesner}. Figure \ref{domedome}(e) shows the time evolution of the central density $|\psi_{i} (r=0,z=0)|^{2}$. After the instability has grown, the oscillations with a large period and large amplitude appear as shown in Fig. \ref{domedome}(e), representing the entire exchange of the central density of $\psi_{1}$ and $\psi_{2}$. 

The growth time of the MI may be determined by the fastest growth mode which has the most negative value of $\Omega_{-}^{2}$.  A straightforward calculation yields $\Omega_{-f}^{2} = -\hbar^{2} k_{f}^{4}/4m^{2}$, where $k_{f}=k_{c}/\sqrt{2}$ represents the wave number of the fastest growth mode. To estimate the growth time $\tau_{f}$, we assume that the density profile of the initial condensate $| 1 \rangle$ has the Thomas-Fermi profile $n_{0}(r,z)=|\psi_{0}|^{2}=\mu_{1}\{1-(r/R_{r})^{2}-(z/R_{z})^{2}\}/g_{1}$ with $\mu_{1}=(15 g_{1} \lambda N / 16 \sqrt{2} \pi)^{2/5} (m \omega^{2})^{3/5}$, $R_{r}=\sqrt{2\mu_{1}/m\omega_{\perp}}=3.9$ $\mu$m and $R_{z}=R_{r}/\lambda=273$ $\mu$m. Because half of the condensate $| 1 \rangle$ is suddenly transferred to $| 2 \rangle$, we use the density $n_{0}(0,0)$ with $N/2$ for $n_{i0}$ ($i=1,2$) in Eq. (\ref{modufreq}). For $N=2\times10^{6}$ we obtain $\tau_{f}=2\pi/|\Omega_{-f}|=26$ msec and $k_{f}=0.42$ $\mu$m$^{-1}$. The value of $k_{f}$ corresponds to the domain size $\ell_{d}=2\pi/k_{f}=15$ $\mu$m. Because of $\ell_{d}>R_{r}$ the instability does not occur in the radial direction. The estimated growth time of 26 msec and the domain size of 15 $\mu$m are in reasonable agreement with the numerical results as well as the experimental results. 

\begin{figure}
\includegraphics[height=0.165\textheight]{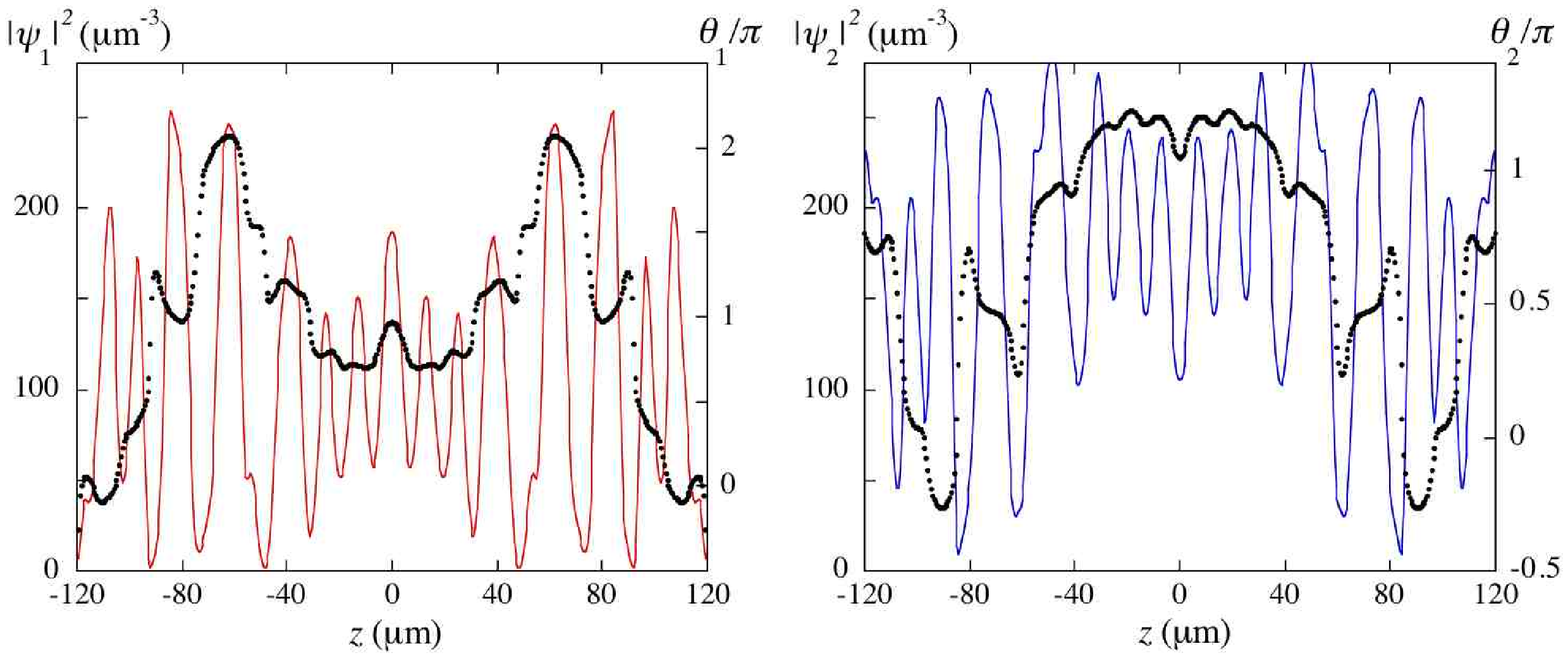}
\caption{The density profile (solid curves) and phase profile (dots) for the multi-domain structure of $\psi_{1}$ and $\psi_{2}$ at $t=225$ msec in the simulation of Fig. \ref{domedome}.}
\label{domepha}
\end{figure}

It is interesting to point out that the dynamics described here is analogous with the collapse dynamics and soliton-train formation in a BEC with an attractive interaction \cite{Saito,Kawaja,Salasnich,Kamchatnov,Carr}. Since the two-component condensates of $^{23}$Na atoms have $g_{1}=g_{12} \equiv g$, Eq. (\ref{2tgpe}) can be rewritten as $i \hbar \partial \psi_{1}/\partial t = ( -\hbar^{2} \nabla^{2} /2m + V_{1} + g n_{T} ) \psi_{1} $ and $ i \hbar\partial \psi_{2}/\partial t = ( -\hbar^{2}\nabla^{2}/2m + V_{2} + g n_{T}  + (g_{2}-g) |\psi_{2}|^{2} )  \psi_{2}$. Here, the total density $n_{T}=|\psi_{1}|^{2}+|\psi_{2}|^{2}$ functions as the static potential because it hardly changes during the time evolution as seen in Fig. \ref{domedome}. Then, an effective attractive interaction appears for the condensate $|2\rangle$ because $g_{2}-g<0$ \cite{Goldstein,Mueller}. This means that the sudden population transfer from $|1\rangle$ to $|2\rangle$ is formally equivalent to the sudden change of the atomic interaction of $| 2 \rangle$ from positive to negative. Therefore, the spontaneously formed domains may have a solitary wave structure like that of a bright soliton train \cite{Kawaja,Salasnich,Kamchatnov,Carr}. This is clearly seen in Fig. \ref{domepha}, where a spatial distribution of the condensate phase $\theta_{i}=$arg$\psi_{i}$ is plotted for a typical multi-domain structure. We can see that the the value of the phase of one component is almost flat within each domain and jumps across the density dips where the domains of the other exist.  Also, there is a similarity to the numerical simulation of bright soliton formation in which bright solitons first form at the sharp edge of an initially prepared condensate with a rectangular shape \cite{Salasnich,Kamchatnov,Carr}. This is explained by the fact that the wavelength (amplitude) of self-interference fringes in the initial wave function is longer (larger) at the edge of the condensate than that at the central part \cite{Carr}. These fringes can be the seed of the modulation, which gets the unstable wavelength $\ell_{d}$ first at the edge. In our simulation, such sharp edges are created {\it spontaneously}, as seen in Fig. \ref{domedome}(b) and (c), because $a_{1}>a_{2}$ and the density of $\psi_{1}$ is thus pushed aside by the repulsion with that of $\psi_{2}$. As a result, the density of $\psi_{1}$ broadens along the axial direction and develops a nearly rectangular shape. 

After the multiple domains have formed, the density in each domain exhibits the complex dynamics. As expected for a bright soliton train in a single-component BEC, the dynamics may be controlled by the effective interaction between the solitary waves, which depends on their phase difference \cite{Kawaja,Salasnich,Carr}. In our simulation, the phase difference between domains shown in Fig. \ref{domepha} is determined nontrivially, following the nonlinear dynamics caused by the MI. Then, an exchange of particles between two domains of the same component occurs as in the Josephson effect, where the domain of the other component has the role of a potential barrier \cite{Ao2}. The oscillation with a large period and large amplitude shown in Fig. \ref{domedome}(e) is the cooperative oscillation of the two-component soliton trains by this Josephson effect. 

\begin{figure}
\includegraphics[height=0.28\textheight]{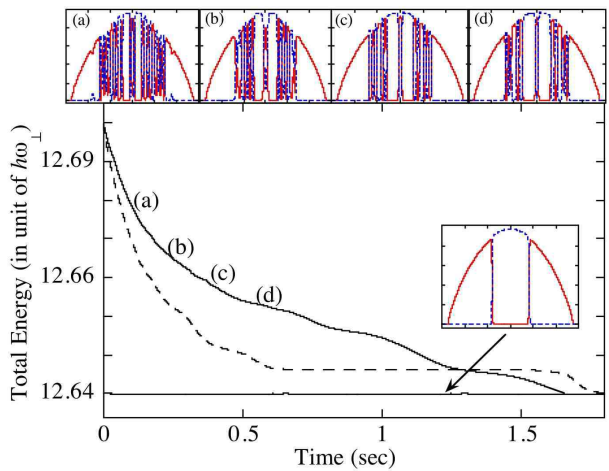}
\caption{The decay of the total energy in the numerical simulation of Eq. (\ref{2tgpe}) with the phenomenological dissipation $\gamma=0.025$ (solid-line) and $\gamma=0.05$ (dashed-line) starting from the multiple domain state of Fig. \ref{domedome} (d). Snapshots of the condensate densities for $\gamma=0.025$ are shown at times (a) - (d). Inset on right shows the densities of the ground state with $E=12.64\hbar \omega_{\perp}$}
\label{domedecay}
\end{figure}

Miesner {\it et al.} observed that the multiple domain structure persisted for at least 20 seconds \cite{Miesner}, which suggests that the domain state is relaxed into a metastable state by some dissipative process. The reason of the metastability was argued to be the large repulsive mean-field interaction energy between the two components and the large kinetic energy that prevents the wave function from speading out in the tight radial direction. To show the metastability, we integrate Eq. (\ref{2tgpe}) from the developed domain state of Fig. \ref{domedome}(d) by introducing a phenomenological dissipation $\gamma$ in the left-hand side of Eq. (\ref{2tgpe}) as $i\partial/\partial t \rightarrow (i-\gamma)\partial/\partial t$ ($\gamma \sim {\cal O}(10^{-2})$  \cite{Tsubota}) under the constraint of a fixed particle number. If a metastable minimum supports the multiple domains, this structure must survive even under dissipation. Figure \ref{domedecay} shows the decay of the total energy $E=\int d {\bf r} \sum_{i=1,2} \psi_{i}^{\ast} (\hbar^{2} \nabla^{2}/2m+V_{i}+g_{i}|\psi_{i}|^{2}/2)\psi_{i}+g_{12}|\psi_{1}|^{2}|\psi_{2}|^{2}$ and some snapshots of the density profile. The simulation shows that the number of domains gradually decreases with a monotonic decrease of $E$, and also that the system certainly decays to the equilibrium state in which one component occupies the middle of the trap and the other sandwiches it \cite{Kasamatsu}. The decay time of the domains is very long because the solitons must move to the condensate edges to vanish. This result indicates that the multi-domain state seen in Fig. \ref{domedome} is not metastable, but instead shows that the solitonic character of the multiple domains can produce a long lifetime; it is generally known that the decay of the topological defects takes a long relaxation time. Although we do not know the specific value of $\gamma$ \cite{Gamma} for Miesner {\it et al.}'s experiment, the experimental condition had negligible thermal components and thus the above solitonic character can explain the extremely long observed lifetime of the multi-domain structure.

In conclusion, we have revealed the nonlinear dynamics of multiple domain formation in two-component BECs and obtained the excellent agreement with the experimental results in Ref. \cite{Miesner}. The underlying mechanism of the domain formation is MI induced by cross-phase modulation, which occurs in condensates with repulsive nonlinearity. This study gives a new interpretation of the observed multiple domains, which are two-component soliton trains and whose long lifetime is ensured by the phase constraint inherent in solitons. Because the presence of MI and solitons are interrelated, our results are closely related with the soliton excitation in multicomponent BECs \cite{Busch} and may be important for the development of nonlinear atom-optics.



\begin{thebibliography}{99}
\bibitem{Cross}
M.C. Cross and P.C. Hohenberg, Rev. Mod. Phys. {\bf 65}, 851 (1993). 
\bibitem{Agrawalbook}
G.P. Agrawal, {\it Nonlinear Fiber Optics} (Academic Press, San Diego, 1995), 2nd ed. 
\bibitem{Donley}
E.A. Donley {\it et al}., Nature (London), {\bf 412}, 295 (2001). 
\bibitem{Strecker}
K.E. Strecker {\it et al}., Nature (London) {\bf 417}, 150 (2002). 
\bibitem{Inouye}
S. Inouye {\it et al}., Nature (London) {\bf 392} 151 (1998).
\bibitem{Saito}
H. Saito and M. Ueda, Phys. Rev. Lett. {\bf 86}, 1406 (2001), Phys. Rev. A. {\bf 65}, 033624 (2002); S.K. Adhikari, {\it ibid}. {\bf 66}, 013611 (2002). 
\bibitem{Kawaja}
U. Al Khawaja {\it et al}., Phys. Rev. Lett. {\bf 89}, 200404 (2002).
\bibitem{Salasnich}
L. Salasnich {\it et al}., Phys. Rev. Lett. {\bf 91}, 080405 (2003).
\bibitem{Kamchatnov}
A.M. Kamchatnov {\it et al}., Phys. Lett. A {\bf 319}, 406 (2003). 
\bibitem{Carr} 
L.D. Carr and J. Brand, Phys. Rev. Lett. {\bf 92}, 040401 (2004). 
\bibitem{Agrawal}
G.P. Agrawal, Phys. Rev. Lett. {\bf 59}, 880 (1987); G.P. Agrawal {\it et al}., Phys Rev. A {\bf 39}, 3406 (1989).
\bibitem{Goldstein}
E.V. Goldstein and P. Meystre, Phys. Rev. A {\bf 55}, 2935 (1997).
\bibitem{Miesner}
H.-J. Miesner {\it et al}., Phys. Rev. Lett. {\bf 82}, 2228 (1999). 
\bibitem{Tim}
E. Timmermans, Phys. Rev. Lett. {\bf 81}, 5718 (1998); P. Ao and S.T. Chui, Phys. Rev. A {\bf 58}, 4836 (1998). 
\bibitem{Stenger}
J. Stenger {\it et al}., Nature (London) {\bf 396}, 345 (1998). 
\bibitem{Burke} 
J.P. Burke, Jr. {\it et al}., Phys. Rev. Lett. {\bf 81}, 3355 (1998). 
\bibitem{Mueller}
E. J. Mueller and G. Baym, Phys. Rev. A {\bf 62}, 053605 (2000); 
M. Ueda, {\it ibid}. {\bf 63}, 013601 (2000). 
\bibitem{Ao2}
P. Ao and S.T. Chui, J. Phys. B {\bf 33}, 535 (2000). 
\bibitem{Pu}
H. Pu {\it et al}., Phys. Rev. A {\bf 60}, 1463 (1999). 
\bibitem{Robins}
N.P. Robins {\it et al}., Phys. Rev. A {\bf 64}, 021601(R) (2001). This paper showed that cross-phase MI does not occur for an antiferromagnetic spinor condensate, e.g., $^{23}$Na atoms. They made MI analysis by taking all three components $|m_{F}=0,\pm1 \rangle$ of the condensate. However, the $|m_{F}=-1 \rangle$ component has been neglected in Ref. \cite{Miesner} as well as our study. This leads to the different conclusion from ours. 
\bibitem{Chui}
S.T. Chui {\it et al}., Physica B {\bf 329-333}, 36 (2003). 
\bibitem{Tsubota}
M. Tsubota {\it et al}., Phys. Rev. A {\bf 65}, 023603 (2002); A.A. Penckwitt {\it et al.,} Phys. Rev. Lett. {\bf 89}, 260402 (2002).
\bibitem{Kasamatsu}
Another equilibrium configuration consists of two phase-separated domains occupying either side of an elongated trap with a domain wall in the middle. Numerical analysis shows that, with the parameter values in Ref. \cite{Miesner}, this configuration is unstable whereas that shown in Fig. \ref{domedecay}(d) is the true ground state. Generally, the stability of the two configurations depends on the trap aspect ratio and the particle number. More details are in K. Kasamatsu {\it et al}., Phys. Rev. A {\bf 64}, 053605 (2001). 
\bibitem{Gamma}
The small change in the value of $\gamma$ changes only the rate that the total energy decreases, as shown in Fig. \ref{domedecay}.
\bibitem{Busch}
P. \"{O}hberg and L. Santos, Phys. Rev. Lett. {\bf 86}, 2918 (2001); Th. Busch and J.R. Anglin, {\it ibid}. {\bf 87}, 010401 (2001); S. Coen and M. Haelterman, {\it ibid}. {\bf 87}, 140401 (2001)
\end{thebibliography}
\end{document}